\documentclass[sn-mathphys,Numbered]{sn-jnl}

\usepackage{graphicx}%
\usepackage{multirow}%
\usepackage{amsmath,amssymb,amsfonts}%
\usepackage{amsthm}%
\usepackage{mathrsfs}%
\usepackage[title]{appendix}%
\usepackage{xcolor}%
\usepackage{textcomp}%
\usepackage{manyfoot}%
\usepackage{booktabs}%
\usepackage{algorithm}%
\usepackage{algorithmicx}%
\usepackage{algpseudocode}%
\usepackage{listings}%

\begin{document}

\title[Article Title]{Metal-Optic Nanophotonic Modulators in Standard CMOS Technology}


\author*[1]{\fnm{Mohamed} \sur{ElKabbash}}\email{melkabba@mit.edu}
\equalcont{These authors contributed equally to this work.}

\author*[1]{\fnm{Sivan} \sur{Trajtenberg-Mills}}\email{sivantra@mit.edu}
\equalcont{These authors contributed equally to this work.}

\author[1]{\fnm{Isaac} \sur{Harris}}
\author[1]{\fnm{Saumil} \sur{Bandyopadhyay}}
\author[1]{\fnm{Mohamed I} \sur{Ibrahim}}
\author[1]{\fnm{Archer} \sur{Wang}}
\author[1]{\fnm{Xibi} \sur{Chen}}
\author[1]{\fnm{Cole} \sur{Brabec}}
\author[1]{\fnm{Hasan Z.} \sur{Yildiz}}
\author[1]{\fnm{Ruonan} \sur{Han}}
\author*[1]{\fnm{Dirk} \sur{Englund}}\email{englund@mit.edu}

\affil*[1]{\orgdiv{Research Laboratory of Electronics}, \orgname{Massachusetts Institute of Technology}, \orgaddress{\street{50 Vassar St}, \city{Cambridge}, \postcode{02139}, \state{MA}, \country{USA}}}


\abstract{Integrating nanophotonics with electronics promises revolutionary applications ranging from light detection and ranging (LiDAR) to holographic displays. Although semiconductor manufacturing of nanophotonics in Silicon Photonic foundries is maturing, realizing active nanophotonics in the ubiquitous bulk CMOS processes remains challenging. We introduce a fabless approach to embed active nanophotonics in bulk CMOS by co-designing the back-end-of-line metal layers for optical functionality. Without changing any of the design rules imposed by a 65 nm CMOS process, we realize plasmonic liquid crystal modulators that exhibit switching speeds 100 times faster than commercial technologies. Our approach, which embeds 'zero-change' nanophotonics into the most ubiquitous platform for integrated electronics, democratizes fabrication of metal-optic nanophotonics, opens the path to mass production of active nanophotonic components, and overcomes major packaging challenges that have previously hindered the realization of complex metal-optic optoelectronic systems.}




\maketitle

\section{Introduction}\label{sec1}

Nanophotonic devices are redefining the technology landscape across numerous sectors, including telecommunications\cite{communication_plasmonics}, sensing\cite{sensing_1}, classical and quantum computing \cite{computing_plasmonics, metamaterial_computing, quantum_computing_photonics}, bioengineering, and renewable energy\cite{energy_me_1, energy_me_2, cooling_2}, enabled by rapid advances in specialized optoelectronic processes such as silicon photonics \cite{saumil}, silicon nitride photonics \cite{adrian}, and lithium niobate photonics\cite{ian}. These developments, particularly in the silicon photonics area, were enabled through the development of the fabless photonics model. The fabless model, which originated in electronic integrated circuits, has significantly impacted the growth and commercialization of photonic integrated circuits\cite{fabless_photonics}. By adopting this model, silicon photonics companies and researchers have leveraged existing silicon-based semiconductor manufacturing infrastructure and expertise, reducing costs and risks associated with building their own fabrication facilities and overcoming the limitations of scarce nanofabrication resources in research institutions. Fabless photonics has thereby accelerated the development and adoption of silicon photonic devices in applications such as data centers, telecommunications, and sensing.

\begin{figure*}%
\centering
\includegraphics[width=124mm]{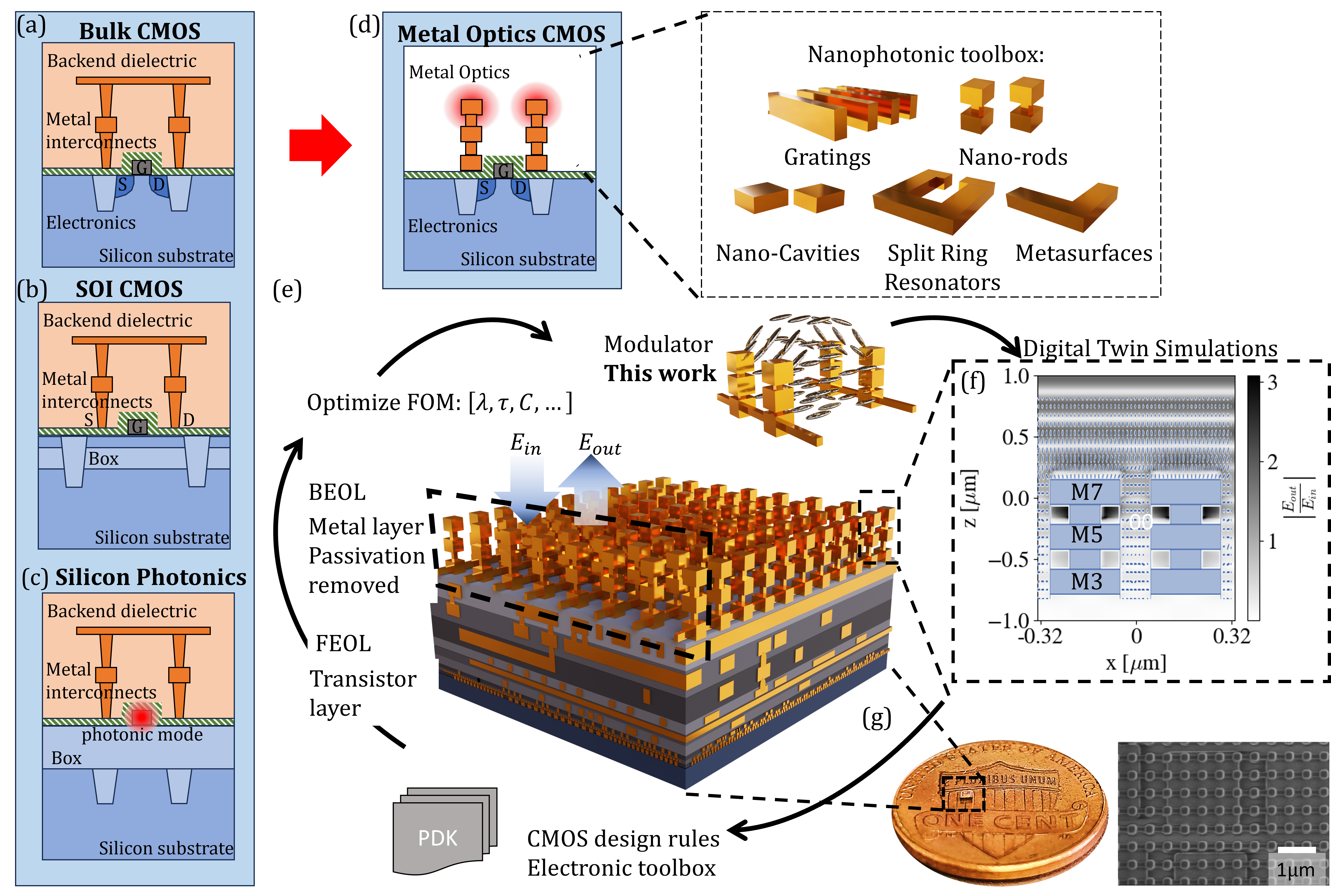}
\caption{\textbf{Metal Optic in bulk CMOS concept} Current standard foundry fabrication methods: (a) Bulk CMOS, where the electronics (in silicon) in the front end of the line (FEOL) are connected through metal wires in the back end of the line (BEOL). The metal layers are surrounded by a dielectric. (b) Silicon-on-insulator (SOI), Where a thin layer of buried oxides ("box") separates the thin silicon top layer from the substrate. (c) Silicon photonics, where a thick layer of buried oxide ("box") separates the dielectric and silicon layer, allowing the support of a photonic mode. (d) Our proposal: Using the Bulk CMOS process, a few metal layers in the BEOL can be repurposed to create metal-optic or plasmonic devices that are compatible with bulk CMOS foundry processes. Without change to the design rules, the BEOL metal layers can be used to create a variety of elements from the nano-photonic toolbox, such as 1D gratings, nano-rods, nano-cavities, split ring resonators, and metasurfaces, and more (inset). Here we design an optical LC based modulator as an addition to this toolbox. This allows for direct integration with the electronic in the FEOL. (e) Co-design of Metal-Optics in CMOS Foundry Processes. We optimize the desired figures of merit (FOM), such as wavelength $\lambda$, rise time $\tau$, or contrast $C$, while adhering to the foundry's design rules. This involves an iterative process wherein the optical, electronic, and mechanical responses of the device are simulated and refined to achieve the targeted FOMs. Subsequently, these are compared against the foundry's design rule check (DRC) to ensure compliance and to verify that they meet the necessary requirements. (g) Image of the fabricated chip on top of a USA penny (left) and SEM image of the nanorods (right).
 }\label{fig1}
\end{figure*}

Despite this progress, an open challenge is the lack of active nanophotonic components in the majority of current semiconductor foundry processes, specifically in CMOS processes. In the bulk CMOS process, shown schematically in Fig. \ref{fig1}(a), the electronics and transistors in the front-end-of-the-line (FEOL) silicon layer are wired through metal layers in the back-end-of-the-line (BEOL), which are supported by a dielectric passivisation material. In silicon-on-insulator (SOI) processes, a layer of buried oxides ("box") separates the device silicon layer from the bulk silicon substrate, Fig. \ref{fig1}(b). Since this is a thin layer, the top silicon cannot support a photonic mode which leaks into the bulk. Silicon Photonics, in Fig. \ref{fig1}(c), has a thick "box" layer, allowing for the support of a photonic mode, but is therefore further away from the electronic device layer, making the integration with electronics difficult. In order to overcome the difficulty of co-integrating photonics with electronics in current processes, researchers have proposed specialized foundry processes (that have very limited commercial availability, and are therefore incredibly expensive), as well as methods requiring heavy post processing.
Co-integration of photonics and electronics in CMOS foundries was demonstrated in silicon-on-insulator processes \cite{ rajeev_2,rajeev_3,rajeev_4, rajeev_undercut}. In addition, specialized foundry processes were customized to create a local photonics region through etching regions in the silicon substrate and depositing an oxide fill and a polycrystalline silicon thin film \cite{rajeev_1}. However, the co-integration of photonic and electronic devices into bulk silicon CMOS foundry processes, which is by far the cheapest, has remained an outstanding challenge. Incorporating nanophotonic devices into standard bulk CMOS processes would be transformative, enabling the integration of photonic devices onto existing CMOS chips. This integration would enable numerous applications, such as LiDAR scanners with CMOS focal plane arrays, hyperspectral biosensing \cite{hatice}, and holographically programmable light-field imagers and displays\cite{Miller_Attojoule}. 

The challenge of co-integrating nanophotonics with electronics within bulk CMOS foundry processes primarily stems from the inability of the bulk silicon layer to support photonic modes (Fig.\ref{fig1} a). Moreover, thin-body fully depleted silicon-on-insulator (SOI) CMOS processes cannot support photonic modes in the thin ($\approx$ 20 nm) crystalline silicon layer \cite{rajeev_1}(Fig.\ref{fig1} b). While SOI substrates have been the primary platform for the monolithic integration of photonics with CMOS (Fig.\ref{fig1} c), they are costly and have a limited supply chain, making them less ideal for high-volume and cost-sensitive applications. Furthermore, photonic SOI processes suffer from limitations due to thermal crosstalk and limited heat dissipation, resulting from the thicker buried oxide layer.
However, a comprehensive examination of the entire CMOS chip structure (as depicted in Fig. \ref{fig1}) reveals that the back-end-of-the-line (BEOL) consists of multiple metal layers that serve as electrical interconnects for the semiconductor devices housed within the FEOL. These metal layers can be patterned with nano-scale resolution. Accordingly, the BEOL can be repurposed to create metal-optic or plasmonic devices that are compatible with bulk CMOS foundry processes. 

In this work, we introduce a fabless model for metal-optics using commercial bulk CMOS foundry processes with zero change to the process design rules. Through careful design of the metal interconnect layers in CMOS chips, it is possible to realize most of the building blocks of metal-optic nanophotonics within the constraints of the foundry design rules (see Fig. \ref{fig1}). To demonstrate the potential of the fabless nanophotonics model, we co-designed several of the metal interconnect layers in 65-nanometer-transistor bulk CMOS process technology to implement both electronic and optical functionalities. Taking advantage of the three-dimensional nanofabrication capability of CMOS foundries, individual plasmonic nanoresonators are electrically interconnected to construct an active plasmonic device. Through minor post-processing, we introduce liquid crystal (LC) molecules to the plasmonic device and realize an on-chip electro-optic modulator.Several metal layers in the BEOL are designed to have multiple functionalities simultaneously: they function as alignment layers for LC molecules, as an optical resonator, and as interdigitated electrodes for the modulator.  Commercial LC optical modulators exhibit a relatively long response time due to the time required for the molecules to reorient themselves under the influence of an electric field. We show that by leveraging the nanoscale resolution of the CMOS metal layers, we can realize smaller effective cell thicknesses that increase the response speed to $> 40 $ KHz, which is two orders of magnitude faster than modern high-speed LC modulators with speeds $< 1$ KHz \cite{fast_LC_1,fast_LC_2}. The proposed CMOS nanophotonic high-speed LC modulator can find applications in telecommunications, reconfigurable optical add-drop multiplexing, optical and quantum computing, sensing and imaging, holographic displays, LiDARs, and miniaturized spectroscopy.

\section{Fabless CMOS Metal-optics Approach}\label{sec2}

Fig. \ref{fig1}e schematically presents the proposed fabless nanophotonics approach using the BEOL of CMOS chips. A few metal layers in the BEOL of a CMOS chip are optimized to meet desired figures of merit (FOM) for their optical response, such as linear or circular retardance, photonic resonance, etc., while satisfying the design rules imposed by the CMOS foundry. In the context of optical modulators, the FOM can include the operation wavelength $\lambda$, the response time $\tau$, the modulation contrast $C$ or others. Following the design rules, a library of nanophotonic components such as gratings, nano-antennas, nano-rods, metasurfaces, ring resonators, etc., can be created. Not all geometries satisfy the design rules of a given CMOS process. For example, some patterns generated through inverse design will not satisfy the foundry design rules \cite{inverse_design}. The remaining metal layers are spared for wiring or connection to different electronic devices in the FEOL, such as transistors for optoelectronic control, photodiodes, CMOS photodetectors, heaters, or electromagnets. Consequently, the proposed approach can integrate nanophotonic devices with advanced electronic devices. 

Depending on the foundry process used, the spatial resolution and the type of metal would differ, with general tendency of advanced processes to have more metal layers and higher spatial resolution. Due to its desirable electronic properties, wires are typically made of copper. Copper is a good plasmonic metal at the red part of the spectrum and has excellent performance in the NIR and IR spectral range \cite{copper}.  Aluminum is used in older CMOS processes and in the top metal layers of most CMOS processes (the ). Aluminum is the best plasmonic metal in the UV range and has good performance across the visible spectrum. It is nearly indistinguishable in terms of optical losses from gold or silver in the NIR range \cite{Al}. 
Depending on the foundry process used, the spatial resolution and the type of metal would differ, with general tendency of advanced processes to have more metal layers and higher spatial resolution. Due to its desirable electronic properties, wires are typically made of copper. Copper is a good plasmonic metal at the red part of the spectrum and has excellent performance in the NIR and IR spectral range \cite{copper}.  Aluminum is used in older CMOS processes and in the top metal layers of most CMOS processes (the "Redistribution Layer" or "Aluminum Pad Layer"). Aluminum is the best plasmonic metal in the UV range and has good performance across the visible spectrum. It is nearly indistinguishable in terms of optical losses from gold or silver in the NIR range \cite{Al}. 

\section{Design of Liquid Crystal Metal-optic Modulator}
Liquid crystal (LC) modulators are optoelectronic devices that control the intensity, phase, or polarization of light through the electro-optical properties of LC molecules \cite{active_plasmonics_LC}. They are stable, inexpensive, and commercially established with applications in displays, communication, adaptive optics, holography, and imaging. Many LC modulators rely on the electric field-induced Freedericksz transition, where LC molecules reorient under the influence of an applied electric field when the applied potential difference $\Delta U$ exceeds the threshold voltage $U_{th}$. The time it takes LC molecules to align with the applied field $\tau$ is given by \cite{PRA_fast}

\begin{equation}\label{eq:tao_rise}
\tau = \frac{\gamma}{\kappa} \cdot \frac{d^2}{\left(\frac{{\Delta U}^2}{U_{th}^2} - 1\right)}
\end{equation}

where $\gamma$ and $\kappa$ are the viscosity coefficient and the elastic constant of the LC, respectively, and $d$ is the LC cell thickness. Fast LC modulators are conventionally phase modulators where the acquired phase depends on the LC optical anisotropy, cell thickness, and operation wavelength. To obtain a desired phase range for a given LC molecule and operation wavelength, the cell thickness is not a design parameter. In the optical range, the cell size is on the order of a few $\mu$m, which results in $\tau > 1$ ms. For many applications such as imaging through scattering media \cite{imagingScattering_roadmap}, video holography \cite{20years}, quantum tomography \cite{tomography}, and quantum control \cite{ian}, the use of high-speed optical modulators is critical. Research on increasing the speed of LC modulators focuses on modifying the mechanical properties of the LC molecules to reduce their viscosity $\gamma$ or threshold voltage $U_{th}$ \cite{fast_LC_1,fast_LC_2} which led to the development of nematic LC modulators with a response time of $ \approx 1 ms$. 

Alternatively, increasing the speed of the LC modulators by decreasing $d$ is possible when modulating a nanophotonic resonance \cite{PRA_fast,LC_metasurface}. We introduce a nanophotonic LC optical modulator that exploits the modulation of a localized plasmon resonance generated in metal nanorods. The modulator design encompasses a two-dimensional array of nanorods that support localized plasmon resonances that depend on the rod height, width, and pitch \cite{Nanorods}. The nanorod array is designed by connecting multiple metal layers (M7, M6, and M5) through vias. The individual nanorods are connected by extending a wire (width$ = 100$ nm) that traverses across them. The wires have three simultaneous purposes: first, they act as a mechanical support for the nanrods; second, they provide an alignment layer for the LC molecules, allowing for the LCs to be drop-casted onto the sample and align along the wires without need for additional alignment layers; and finally, they act as interdegitated electrodes, supplying the potential difference and electric field for the modulator operation.
We simulated the response of LC molecules embedded in the designed plasmonic structure. The LC molecules initially align parallel to the wires; however, When a bias is applied across the electrodes, the LC molecules re-orient along the fringe field which modifies the refractive index sensed by the plasmon resonance. Fig. \ref{fig2}(a) shows a simulation of these dynamics, without the bias (top) and with it (bottom), where a change in the orientation of the liquid crystal molecules (light blue lines) is apparent. 

\begin{figure*}
  \centering
  \includegraphics[width=124mm]{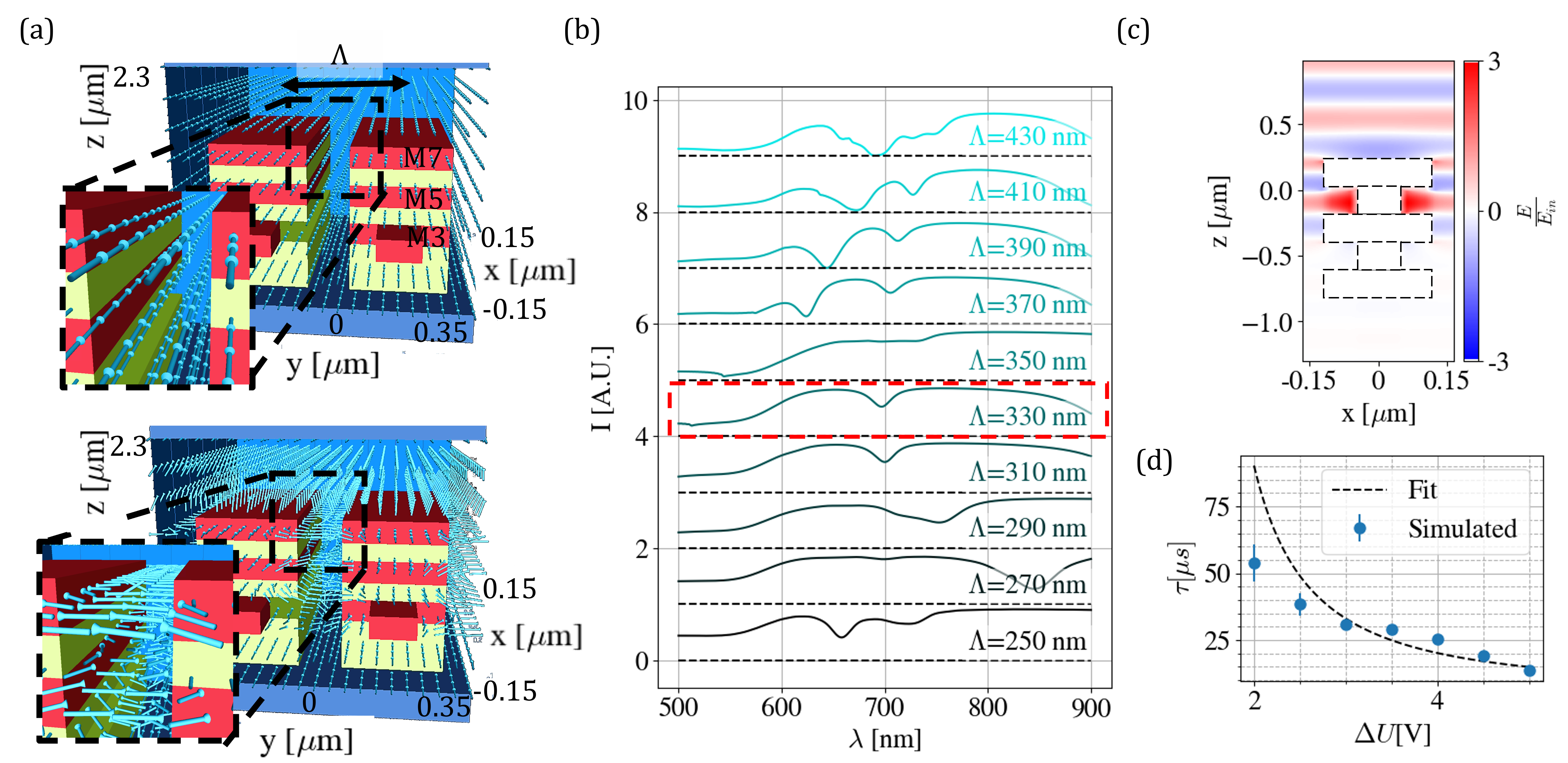}
\caption{\textbf{Design of liquid crystal Nanophotonic modulator with Zero-change to the foundry rules.} (a) Images from LC dynamics simulations where the distance between neighboring units $\Lambda$ is shown. The thickness is determined by the design rules of the interconnect layers. (Top) The liquid crystals align along the trenches which act as an alignment layer. (Bottom) When a potential difference is applied the LC orient along the fringe field lines created by the two electrodes. (b) Simulated reflection spectra as a function of wavelength ($\lambda$) for different pitches ($\Lambda$). Highlighted in red is the chosen design, which complies with our targeted wavelength of 700 nm. (c) Simulation of the unit cell field. (d) Simulated rise time $\tau_{rise}$ per applied voltage for our chosen $\Lambda$.}
\end{figure*}\label{fig2}

Our design targets the modulator to operate at a wavelength of 700 nm, which is of interest to quantum computing technologies \cite{ian}.  We calculated the nano-rod array resonance as a function of the array's pitch (Fig. \ref{fig2}(b)). A nano-rod pitch of $\Lambda =$ 330 nm and a rod width of $w =$ 230 nm exhibits a resonance at 700 nm. The electric field simulation of the nanorods at resonance is shown Fig. \ref{fig2}(C) (see Methods for details on the field simulations). In our simulations, we assumed an LC layer as a superstrate with $n = 1.55$. A field enhancement factor of 3 for the x component of the electric field occurs primarily in the region between the nanorods. The optimized nano-rod array dimensions are then incorporated into simulations of the LC dynamics (see Methods). Fig. \ref{fig2}(d) shows the simulated $\tau$ for different applied voltages. As the bias increaases, $\tau$ decreases and reaches  0.0138$\pm$0.0002 ms at $\Delta U = 5 V$.  From the fit parameters we can extract the effective cell thickness (see Methods) according to Equation \ref{eq:tao_rise}. Based on the simulated $\tau$, the effective cell thickness, $d_{eff}$, is 81 $\pm$ 11 nm.

\section{Fabrication and Experimental Results}

Our designed modulators were fabricated by a CMOS foundry (TSMC) in their 65-nm bulk CMOS process. The passivation layer of the as-received chips was removed through reactive ion etching (see Supplementary Materials). In Fig. \ref{fig3}(a), we delineate the micro- and nano-scale features of the chip. SubFig. \ref{fig3}(a)(i) show the $1 mm \times 1 mm$ chip next to a US cent coin. SubFig. \ref{fig3}(a)(ii) provides a view of the full CMOS chip following the removal of the passivation layer. Using scanning electron microscopy (SEM), SubFig. \ref{fig3}(a)(iii) and \ref{fig3}(a)(iv) depict nanorod devices, with different pitches. In SubFig.e \ref{fig3}(a)(V), shows an SEM image where the wires are visible demonstrating that multilayer nanophotonic structures can be realized using our fabless CMOS metal-optics approach. To create interdigitated electrodes, every other wire in M5 (false color, purple) is terminated and connected through a via (false color, yellow) to an M4 metal layer which is routed to a wire bond pad on the PCB. The remaining wires extend further and are connected to a separate pad. 

\begin{figure*}
  \centering
  \includegraphics[width=134mm]{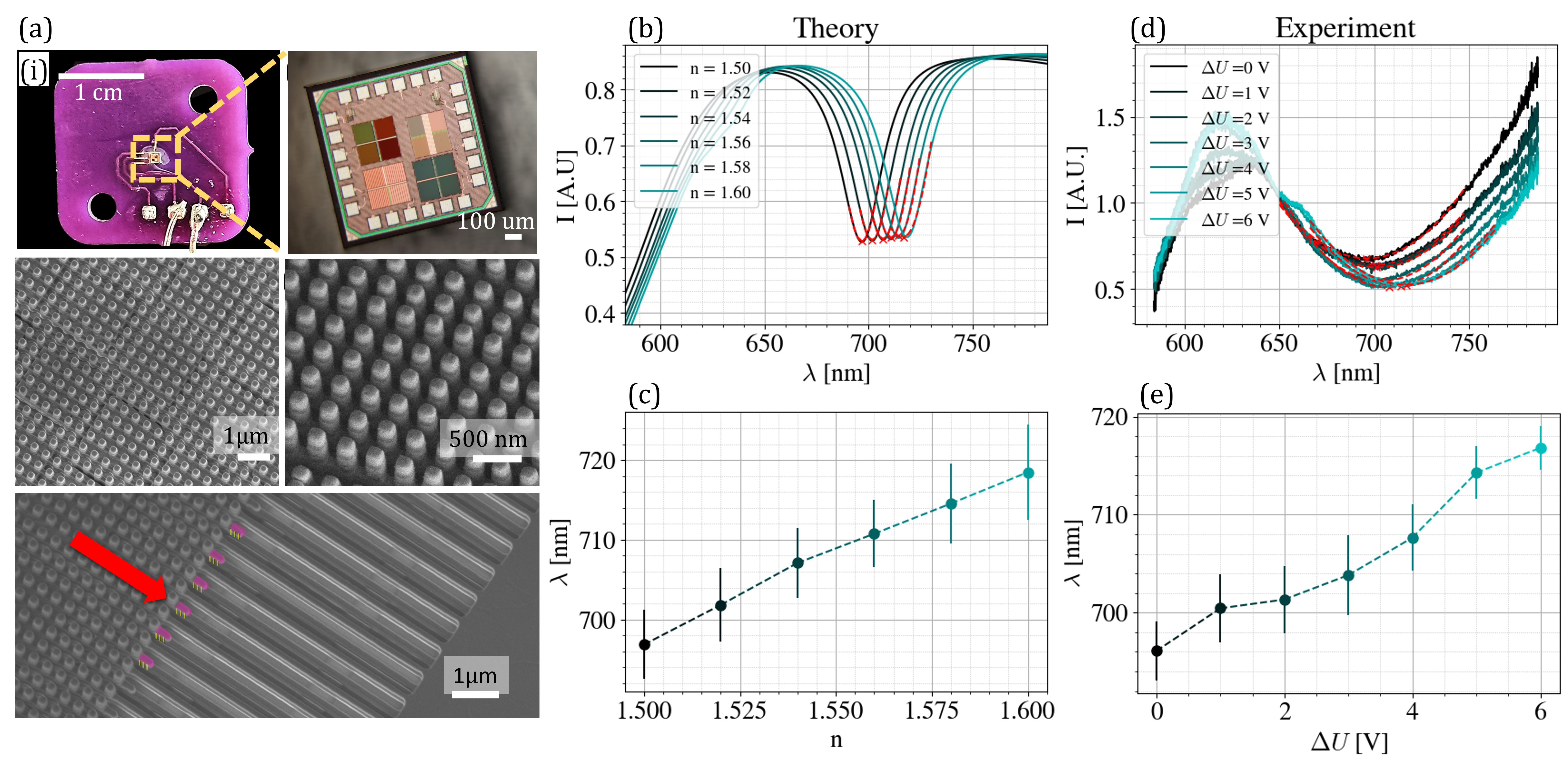}
  \caption{\textbf{Nanorod plasmonic liquid crystal modulator.} (a) Shows images of the chip. (i) Image of the chip wire bonded to a PCB with a liquid crystal drop on top. (ii) Image of the chip after removing the passivation layer. The modulator design is located on the bottom right (green squares). The small squares surrounding the designs are metal pads. (iii-v) SEM images of the 2D nano rods. In (v) the vias and upper wires are colored in yellow and pink (false color). (b) Numerically calculated reflectance vs. wavelength for a change of the refractive index, and (c) the extracted resonant shifts as a function of the change in the ambient refractive index. (d) Experimentally measured reflectance as a function of wavelength as a function of the applied voltage $ \Delta U$, and (e) the extracted resonant shifts per $ \Delta U$.
  }
\end{figure*}\label{fig3}

To realize an LC modulator, we drop-cast LC molecules (E7- Merck) on the device and applied a bias $\Delta U$ across the electrodes. Fig. \ref{fig3}(b) displays the theoretically calculated reflectance spectrum in relation to the refractive index of the superstrate. The extracted minimum reflectance wavelengths, representing the resonant wavelength, are illustrated in Fig. \ref{fig3}(c). The experimentally measured reflectance vs. the $\Delta U$ and the corresponding minimum reflectance wavelength are shown in Figs. 3(d) and Fig. \ref{fig3}(e), respectively (see supplementary materials for details on the experimental setup). The results indicate that as we increase the voltage the plasmonic resonant red-shifts, as expected from LC with positive birefringence.

To measure the modulator response time of our modulator, we applied an AC square wave and measured the device reflectance. Fig. \ref{fig4}(a) and Fig. \ref{fig4}(b) shows the dynamics of the electro-optical response of the plasmonic LC modulator as a function of time, for different values of $\Delta U$. The reflection switching fronts can be approximated with single exponential fits \cite{PRA_fast}(see Methods), while the decay time is expected to remain consistent across varying bias. The measured rise times (decay times) obtained by fitting the reflection switching fronts are shown in Fig. \ref{fig4}(c) (Fig. \ref{fig4}(d)). The values reported are the average of multiple fits from the captured time series at $300$ Hz. The error bars represent the standard error of the mean obtained from the multiple fits and we confirmed these results by measuring the 20-80 rise time shown in Fig. \ref{fig_20-80}, which supports the fit results (see Methods and appendixes). For the decay time, a weighted average was computed and is illustrated in Fig. \ref{fig4}(d) in a black dashed line, with the standard deviation from the mean depicted in blue. The average decay time $\tau_{decay}$ is 30.4 $\pm$ 2.1 $\mu$s.  For rise times, at  $ \Delta U = 5 V$, the measured rise time is $\tau$ = 27.49 $ \pm$ 3.33  $ \mu s$ which corresponds to a modulation speed $36 \pm 4$ KHz, which is two orders of magnitude faster than state-of-the-art LC modulators. 

\begin{figure*}
  \centering
  \includegraphics[width=134mm]{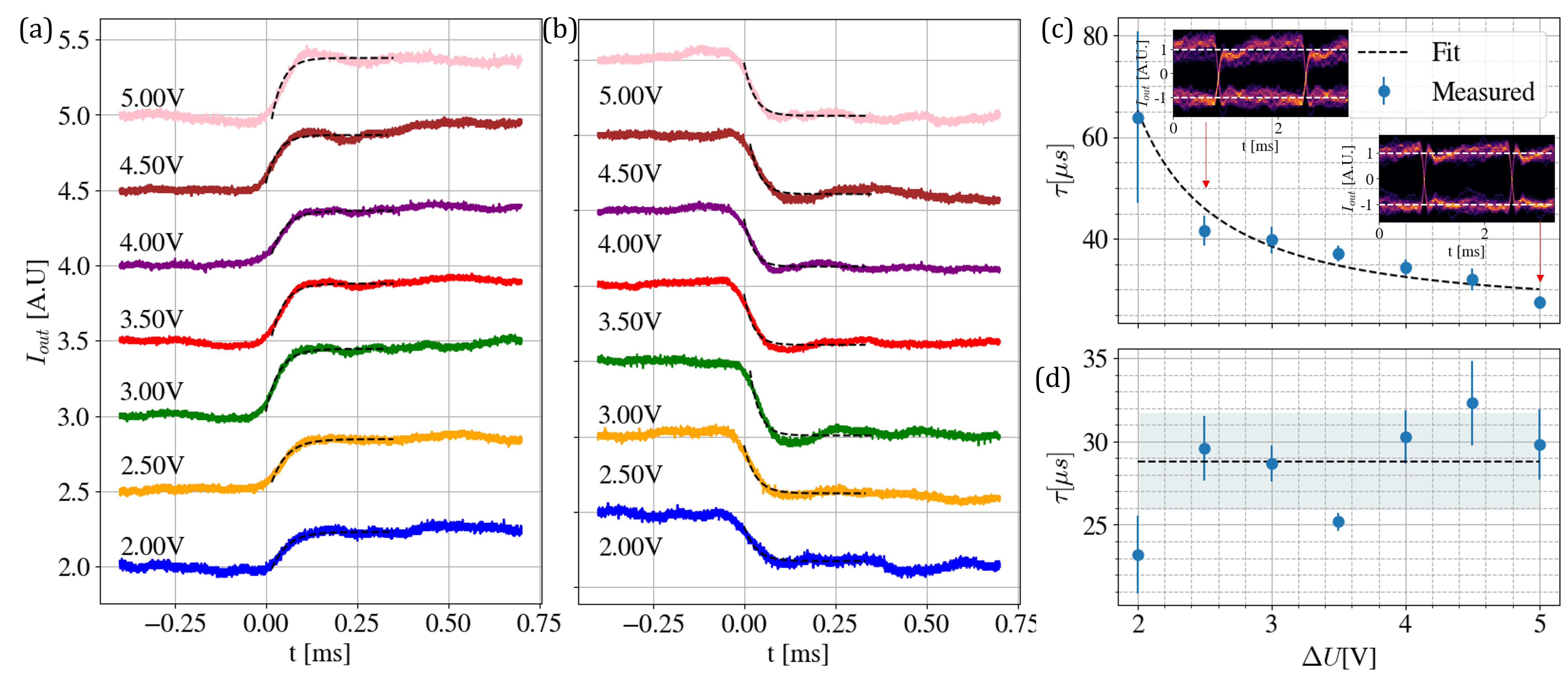}
\caption{\textbf{Modulator electro-optic dynamics} (a) Rise time and (b) decay time. raw data (solid, color) and fit (black, dashed) of intensity vs. time for different $\Delta U$. $\Delta U$ is applied at t=0 sec. (c) Rise time and (d) decay time measured $\tau$ (from fit) vs. $\Delta U$ (blue points) and fit (black dashed line). Blue shaded region show the std of the mean. Inset are eye diagrams at $\Delta U= 2.5 V$ and $\Delta U = 5 V$.}
\end{figure*}\label{fig4}

The rise time vs. voltage results are then fitted with equation \ref{eq:tao_rise} to extract the effective cell size $ d_{eff}$ and the value of $\tau$ at infinite potential difference $ \tau_{infty}$ in Fig. \ref{fig4}(b). For typical parameters of E7 liquid crystals ( viscosity coefficient $ \gamma= 0.190 \ Pa s^{-1}$, elastic constant $ ~ 11 pN $ and $V_{th} ~ 1.3 V$), the calculated $d_{eff}$ is 55.5 $\pm$ 6.9 nm and the free parameter $ \tau_{\infty}$ is $26.2 \pm 3.4 \mu s$. We found that the fitted value of the cell size $d_{eff}$ is smaller than the numerically predicted effective cell thickness, which is likely due to the immobilized liquid crystal molecules near each interface, reducing the effective cell size. A similar calculation can be performed for the decay time, see Appendix \ref{sec:20-80}. 

\section{Conclusions and Future Outlook}

We have introduced a fabless model that enables the fabrication of metal-optic devices and the co-integration of photonics and electronics in bulk CMOS foundry processes. Using a 65-nm CMOS process, we fabricated arrays of metallic nanorods, which are among the most important building blocks in metal-optics due to their strong optical nonlinearities and high refractive index sensitivity  \cite{nanorod_zayats_1, nanorod_zayats_2, nanorod_zayats_3, nanorod_zayats_4}. We expect that future works will develop plasmonic nano-rod-based devices in CMOS foundries and directly integrate them with CMOS electronic devices.

With minimal post-processing, we integrated liquid crystals into the nano-rod array and demonstrated an LC optical modulator with speeds two orders of magnitude faster than those of state-of-the-art LC modulators. This work serves as a foundation for fabricating electro-optic devices in CMOS foundries, including the creation of LiDARs and spatial light modulators. Other electro-optic materials with higher response speeds, such as electro-optic polymers, can also be integrated into these devices.

The fabless nanophotonic model has the potential to spur growth in the metal-optic research and industry, similar to the impact of the fabless model on silicon photonics. Due to fabrication difficulties, many metal-optic devices were demonstrated at microwave frequencies \cite{engheta_1, engheta_2}. Moreover, metal-optic devices that require multilayer fabrication with precise alignment can be easily demonstrated using the proposed metal-optics fabless model \cite{ndao_ep, EIT}. The possibility of integrating metal-optic devices with other electronic components commonly fabricated in CMOS foundries—such as electromagnets, heaters, transistors, and CMOS photodetectors—will pave the way for a new generation of nanophotonic devices.

\section{Methods}\label{sec11}

\subsection{Removing the CMOS passivation layer}
In CMOS chips a passivation layer protects the underlying semiconductor devices and circuitry from damage and degradation. We removed this passivation layer to integrate LC molecules with deep metal layers in the CMOS stack.

The passivation layer is typically made of dielectric materials with good insulating and chemical barrier properties. Common materials used for passivation layers include silicon dioxide ($SiO_2$), silicon nitride ($Si_3N_4$), and silicon oxynitride (SiON). We used reactive ion etching (RIE) , (Plasma-Therm) to remove these layers.  The etching was done using Carbon Tetrafluoride ($CF_4$) gas. Note that the  Cu nanostructures act as a hard mask during the RIE process. This means that right below the Cu structure, a dielectric layer is present. The exact nature and optical properties of the dielectric layer was not supplied by the manufacturer. The etching rate was $~ 40$ nm/min and took a total of 3 hours using $200$ W and $15$ mTorr. The etching process was done in steps of 15 minutes to allow for the sample to cool down and to avoid overheating of the substrate, which would collapse the nanostructures. We found that these conditions vary significantly depending on the tool used.

\subsection{FDTD calculations}
All numerical simulations were carried out using the finite difference time domain (FDTD) method (Lumerical\textsuperscript{™}). When designing our plasmonic devices, we used the perfectly matched layer (PML) boundary condition in $z$ direction and the periodic boundary conditions in the $x$ and $y$ directions. The optical constants of Cu were taken from Palik. The ordinary and extraordinary refractive indices of the liquid crystal molecules (E7-Merck) are 1.55 and 1.75, respectively. The unetched dielectric is assumed to be  silica with a refractive index = 1.45.

\subsection{CMOS Layout}
Following the FDTD modulator design stage, the modulator designs were imported into Cadence Virtuoso.
Here, we confirmed that they followed design rules for the 65 nm process and added connections to underlying metal layers for the interdigitated electrodes.
Different designs were combined onto a single die, and the electrodes were routed to wirebond pads for testing.

\subsection{Reflection measurements}
For our reflection measurements we mounted the sample on a microscope with a broadband source (lamp) and coupled the reflected light to a multimode fiber connected to a Thorlabs spectrometer (CCS200). Since the resonance is very broad, it is sufficient to fit it with a quadratic polynomial in order to estimate the resonant wavelength. We fit a quadratic polynomial around the dip:
\begin{equation}\label{eq:quadratic}
I = \alpha \lambda ^2 + \beta \lambda + C.
\end{equation}
Where $\alpha, \beta$ and $C$ are fitting parameters. The resonant wavelength $\lambda_{min}$ is the minimal value:

\begin{equation}\label{eq:min_wav}
\lambda_{min} = -\frac{\beta}{2\alpha}
\end{equation}

We extract the error for the value from the square root of the covariance matrix of the fit, which gives errors $\Delta\alpha, \Delta\beta$ on the fit parameters, and calculate the error $\Delta \lambda_{min}$ for $\lambda_{min}$: 

\begin{equation}\label{eq:min_wav_error}
\Delta \lambda_{min} = \frac{\beta}{2\alpha^2}\Delta \alpha + \frac{\Delta \beta}{2\alpha}
\end{equation}
The same procedure was applied for measuring the resonant wavelength of the simulated spectra. 

\subsection{Response time measurements}
We built an optical microscopy setup (Fig. \ref{fig_setup}) to measure the response time of our devices. We used a broadband supercontinuum source (SuperK EXTREME Super-continuum Lasers - EXW-12 from NKT Photonics). The source is filtered with a 690 nm filter with a FWHM of 10 nm and the beam is expanded and collimated on our sample using a 50X Mitutoyo IR objective. The reflected light is imaged onto a visible camera (Thorlabs Zelux® 1.6 MP Color CMOS camera) as well as  directed to a photo-detector (Thorlabs APD430C) that connects to an oscilloscope/spectrometer. All rise time measurements were performed on this optical setup.
Time series measurements were acquired from the APD output on an oscilloscope. We used a signal generator to produce a square wave at 300 Hz applied to the sample and applied a high-pass filter on the measured data to remove slow drifting of the entire field. For each clock period the measured intensity was fitted to an exponential:

\begin{equation}\label{eq:rise_time}
I_{out,i}(t) = I_0exp\left(\frac{t-t_0}{\tau}\right)+c
\end{equation}

Where $I_{out,i}$ is the measured intensity of the $i_{th}$ instance, $I_0$ is the amplitude of the fit, $t$ is time, $t_0$ is the zero time, $c$ is a constant shift and $\tau$ is the rise time. The final value was calculated according to the central limit theorem: 
\begin{equation}\label{eq:average}
\tau = \hat{\tau} \pm \frac{\sigma}{\sqrt{N-1}}
\end{equation}
Where $\hat{\tau} = \frac{\Sigma_i \tau_i}{N}$ is the mean value, $\tau_i$ is the fit of the $i_{th}$ instance, $N$ is the number of instances and $\sigma = \sqrt{\Sigma_{i}\frac{(\tau_i - \hat{\tau})^2}{N}} $ is the standard deviation. 

\subsection{Effective Cell Thickness Calculation}
We calculated $\tau$ for each voltage from 2 V to 5 V and fit the data (weighted least squares) to:
\begin{equation}\label{eq:fit_equation}
\tau = \tau_{\infty} + \frac{b}{\frac{U^2}{U_{th}^2}-1}
\end{equation}
where $\tau_{\infty}$ is the minimum value of $\tau$ in the limit of $U\rightarrow \infty$, $U$ is the voltage, $U_{th}$ is the threshold voltage for E7 LC and is taken to be 1.3 V \cite{PRA_fast}, and $b$ = $\frac{\gamma d^2}{\kappa}$, where $\gamma$ and $\kappa$ are the viscosity coefficient
and the elastic constant of the LC, respectively,
and $d$ is the LC cell thickness. The effective cell thickness can then be extracted from the fit:

\begin{equation}\label{eq:d}
d_{\mathrm{eff}} = \sqrt{\frac{b \gamma}{\kappa}} \pm \frac{\Delta b}{2} \sqrt{\frac{\gamma}{b \kappa}} 
\end{equation}

Where $\Delta b$ is the error fir the fit parameter b. For the decay time, the same equation can be used,  $ b \rightarrow \tau_{decay}$.

\subsection{LC dynamics simulations}

We used SHINTECH OPTICS LCDMaster 3D software to model the LC dynamics. The software allows the user to simulate complex 3D rotation of liquid crystals. Fig. \ref{fig1} and Fig. \ref{fig2}(d) in the manuscript show the orientation of LC molecules calculated using the LCDMaster software with a bias OFF and ON, respectively. The dimensions for the device are the same as the one used in FDTD simulations and obtained from TSMC. The unit cell has a height of $2.3~\mu m$ and base dimensions of 2$\Lambda$ and $\Lambda$ (where $\Lambda$ is the pitch). To simulate the liquid crystal dynamics for an infinite 2D array of nanorods, we treat the unit cell with periodic boundary conditions in the $x$ and $y$ directions. The minimum distance between meshes was taken to be $7~nm$. The temporal resolution used to simulate dynamics was 0.08 $\mu s$, and the simulation length was set to be 1 ms. A step voltage is applied across the nanorod electrodes at the start of the simulation (i.e. we set the adjacent nanorods to have voltages of $+\Delta U/2$ and $-\Delta U/2$).

The initial alignment director has a pre-tilt of 2 degrees on floor and ceiling. The anchoring conditions were set to zero anchoring on the ceiling (since we drop cast the LCs and the interface is with air) and weak anchoring on the floor. The simulation parameters used were: $K1 = 11.0~pN, K2 = 10.2~pN, K3 = 16.9~pN$,
$\epsilon_{\parallel}= 19.6$, $\epsilon_{\perp} = 5.1$ \cite{jacobs1992liquid}, and $\gamma = 0.19 ~Pa.s$ \cite{gorkunov2017fast}. After running the simulation, at each timestep we calculated the angular deviation $\Delta \psi$, defined as:
\begin{equation}\label{eq:angular}
\begin{split}
\Delta\psi(x,y,z, t) = \theta(x,y,z,t)-\theta(x,y,z,t_0) + \\
\phi(x,y,z,t)-\phi(x,y,z,t_0)
\end{split}
\end{equation}
where $\theta(x,y,z,t)$ and $\phi(x,y,z,t)$ are the azimuthal and polar angles of the LC at time $t$ as defined in Fig. \ref{fig-sims}(a). The angular deviation was integrated over a volume $V$, corresponding to the region where the field is large (according to FDTD simulations):
\begin{equation}\label{eq:psi}
\psi(t)_{tot} =  \int_V dx dy dz\Delta\psi(x,y,z,t)
\end{equation}

Finally, $\psi(t)_{tot}$ was fitted to an exponential (see equation \ref{eq:rise_time}) to extract $\tau$, as can be seen in Fig. \ref{fig-sims}(a).  

\backmatter

\bmhead{Acknowledgments}
We thank the MITRE Quantum Moonshot Program for program oversight and funding. Experiments  were  supported in  part  by  Army  Research  Office  grant  W911NF-20-1-0084. We thank the MIT.nano staff for fabrication assistance. S.T.M. was supported by the Schmidt Postdoctoral Award and the Israeli Vatat Scholarship. C.B. is supported by NSF GRFP. H.Z.Y. and A.W. were supported by MIT UROP program. I.H. is supported by the STC Center for Integrated Quantum Materials (CIQM) NSF Grant No. DMR-1231319, the National Science Foundation (NSF) Engineering Research Center for Quantum Networks (CQN) awarded under cooperative agreement number 1941583, and the MITRE Moonshot Program. 

\bmhead{ Authors' contributions}
 D.E., R.H., and M.E. conceived the idea. M.E. and S.T.M. led the research and performed the experiments, built the experimental setups and performed the data analysis. M.E. developed the post fabrication procedure, designed the plasmonic device and liquid crystal modulator, and performed the FDTD simulations. I.H.,  M.I., and X.C performed the CMOS tapeout. S.B. designed the printed circuit board (PCB). S.B. and H.Y. helped with the rise time experiments. C.B. contributed to the data analysis. A.W. performed the liquid crystal simulations. D.E. and R.H. supervised the research. M.E., S.T.M. and D.E. wrote the manuscript with input from all authors.

\begin{appendices}

\pagebreak
 
\section{Optical Setup}\label{secA1}

\begin{figure}[h]
  \centering
\includegraphics[width=84mm]{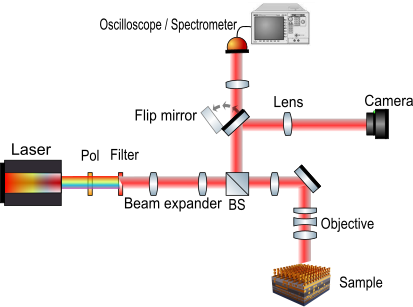}
\caption{ Schematic depiction of the optical measurement setup.  BS - beam splitter, Pol - Polarizer.} \label{fig_setup}
\end{figure}

Illustrated in Fig. \ref{fig_setup} is a schematic representation of the optical configuration employed in our experimental setup. The broad-spectrum source is polarized and is subsequently filtered through a 690 nm bandpass filter. It is then collimated onto the sample utilizing a 4-$f$ system comprising a focusing lens and an objective (Mitutoyo NIR objective). The reflected light can be directed towards a fiber coupled photodetector (Thorlabs APD430), connected to a spectrometer or oscilloscope, or a camera through the use of a flip mirror. The sample is affixed to a printed circuit board (PCB) and is subject to controlled adjustments for tip/tilt. The same setup was used for measuring the rise time and decay time.

\section{Passivation Layer Removal}\label{secA2}

\begin{figure}
  \centering
  \includegraphics[width=74mm]{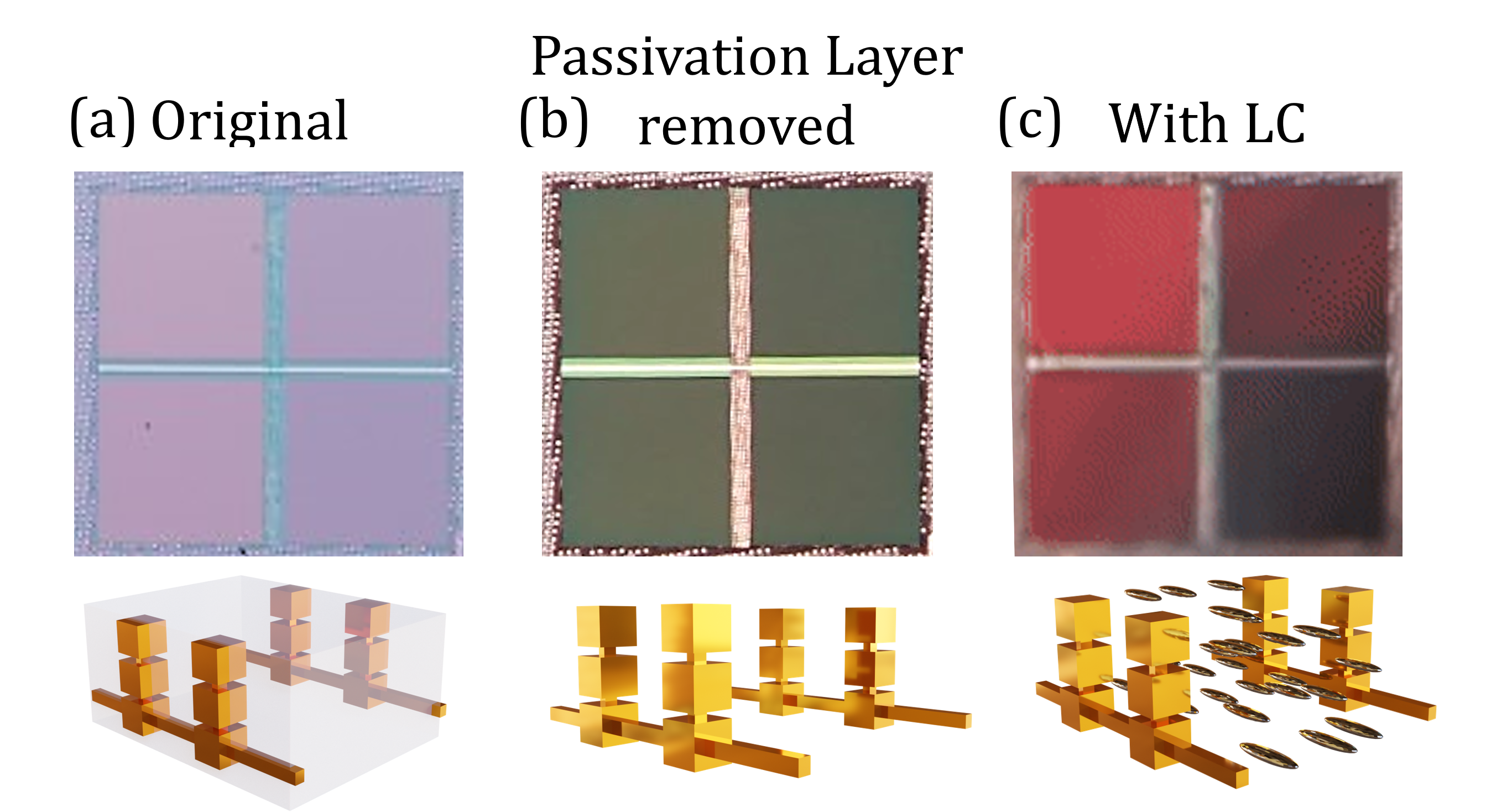}
  \caption{Post-fabrication of CMOS LC modulators: (a) Original CMOS LC modulator before post-processing. (b) CMOS LC modulator after removing the passivation layer. (c) CMOS LC modulator after adding the liquid crystals. The four pixels possess distinct $\Lambda$ values and our investigation focuses on the bottom-left pixel in this study.}\label{fig_fab}
\end{figure}

During the post-processing stage involving the removal of the passivation layer through ion etching, a change in pixel color becomes apparent throughout the procedure. The pixels retaining the passivation layer exhibit a diffused red hue. As depicted in Fig. \ref{fig_fab}, alterations in pixel color are observed subsequent to the removal of the passivation layer and the introduction of LCs. Specifically, the transition ensues from a diffused red tone to green, eventually reverting to a vivid red with the incorporation of LCs. The change in color confirms that the nano-rods support a plasmonic resonance that is sensitive to variations in the local refractive index.

\section{20-80 Rise time analysis.}\label{sec:20-80}
\begin{figure}
  \centering
  \includegraphics[width=144mm]{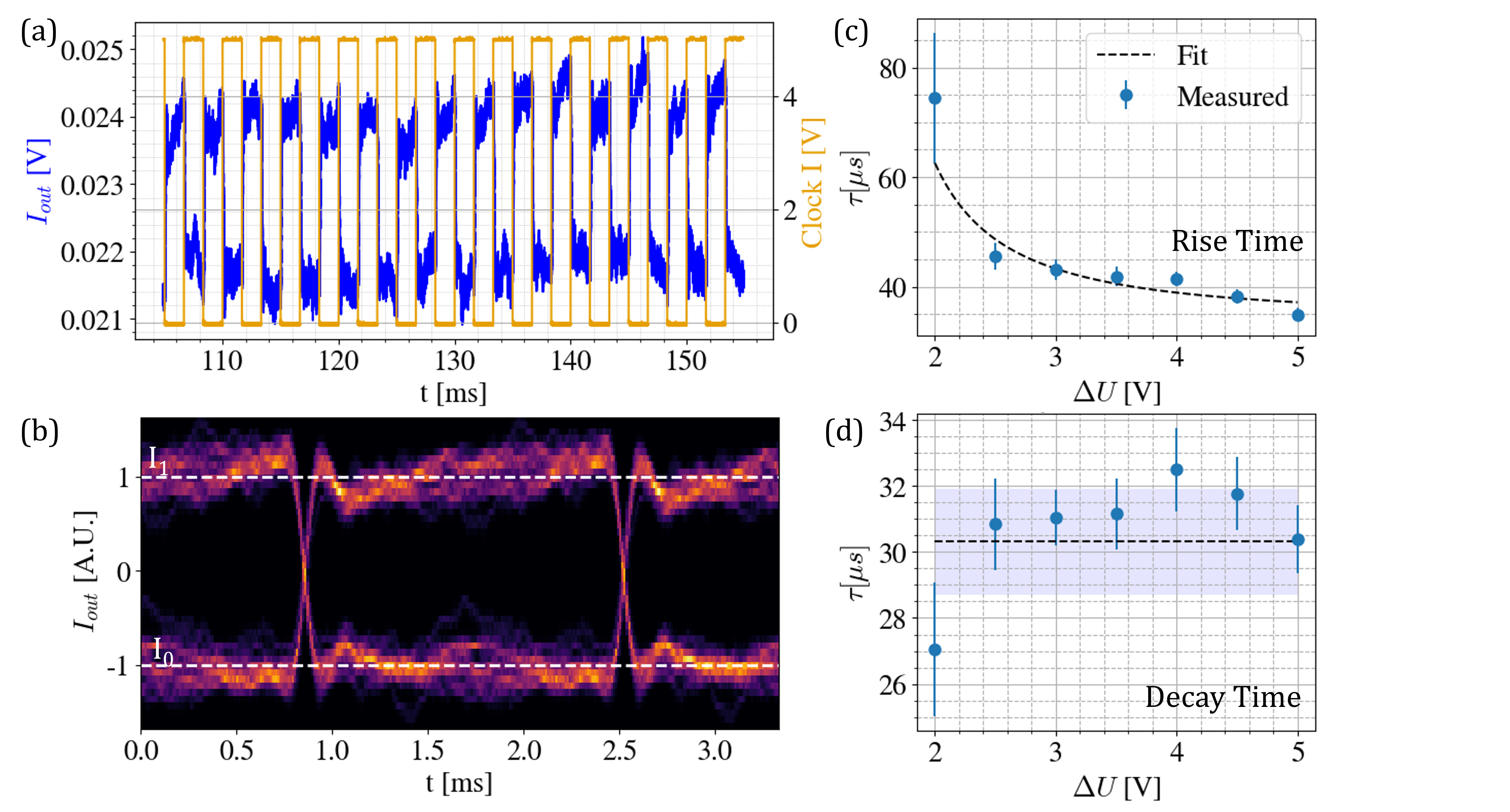}
  \caption{Time series reflection measurements. (a) Reflection measurement  for an applied voltage of 5 V. The raw reflected data (dark blue) and the applied square wave (yellow) can be seen. (b) Eye diagram for $\Delta U$=5 V. (c) Rise time extracted from 20-80 measurement at different voltages (blue dots) and fit (black dashed line). (d) (c) Decay time extracted from 20-80 measurement at different voltages (blue dots) and mean (black dashed line) with standard deviation (blue). }\label{fig_20-80}
\end{figure}

In addition to fitting an exponent to retrieve the rise time, we performed a 20-80 rise time analysis. We constructed an eye diagram from a time-series measurement at 300 Hz for every applied $\Delta U$, as can be seen in Fig. \ref{fig_20-80}(b). The $I_{0}$ and $I_{1}$ values were calculated as the average value of all the data samples captured inside the middle 20\% (40\% to 60\% points) of the eye period. The 20-80 rise time $t_{20-80}$ was calculated as the time difference between the last cross of the data with 20\% of the total height, $I_{20\%}) = 0.2 (I_{1}-I_{0}) + I_{0}$, and the first cross of the data with 80\% of the total height $I_{80\%}) = 0.8 (I_{1}-I_{0}) + I_{0}$.
\begin{equation}\label{eq:I_tau-20-80}
t_{20-80, rise} = t(I = I_{80\%}) - t(I = I_{20\%})
\end{equation}

For the decay time of the modulator $\tau_{decay}$, the analysis was based on equation \ref{eq:I_tau-20-80}, with a modification: the 20-80 decay time is defined as the time difference between the last instance when the data crosses 80 \% of the total height and the first instance when it crosses 20 \%: 

\begin{equation}\label{eq:I_tau-20-80}
t_{20-80, decay} = t(I = I_{20\%}) - t(I = I_{80\%})
\end{equation}

In  both cases, We used equation \ref{eq:average} to calculate the rise time and error for every $\Delta U$. The relation between $t_{20-80}$  (both for rise time and decay time) and $\tau$ is given by:

\begin{equation}\label{eq:tau-20-80}
t_{20-80} = 1.386\tau
\end{equation}

Fig. \ref{fig_20-80}(c) shows a fit to equation \ref{eq:fit_equation}. The effective cell thickness was calculated according to equation \ref{eq:d}. The fitted value is $d_{\mathrm{eff}}$ = 47.2 $\pm$ 6.5 nm and $\tau_{\infty}$ = 34.9 $\pm$ 3.1 $\mu s$. These values are within the error of values reported in the main text. The minimal measured value $\tau$ = 28.18 $\pm$ 2.16  $\mu s$ at $\Delta U$ = 5 V.

Our expectation is that the decay time remains consistent across varying bias. A weighted average of the decay times was computed and is illustrated in Fig. \ref{fig_20-80}(d). The average decay time $\tau_{decay}$ obtained from the 20-80 analysis is 30.7 $\pm$ 1.5 $\mu$s. This value is well within the standard deviation of the value reported from the exponential fit analysis. 

We can derive the effective cell thickness for the decay time from equation \ref{eq:d}. This calculation yields $d_{\mathrm{eff}}$ = 41.9 $\pm$ 1.5 nm and $d_{\mathrm{eff}}$ = 42.1 $\pm$ 1.0 nm for the exponential fitting and 20-80 analysis respectively. The effective cell thickness calculated from the measured decay time and measured rise time are within the error of both, which suggests that the two measurements are statistically consistent within the uncertainties of each method. 

\section{Simulated LC dynamics.}
\begin{figure}
  \centering
  \includegraphics[width=104mm]{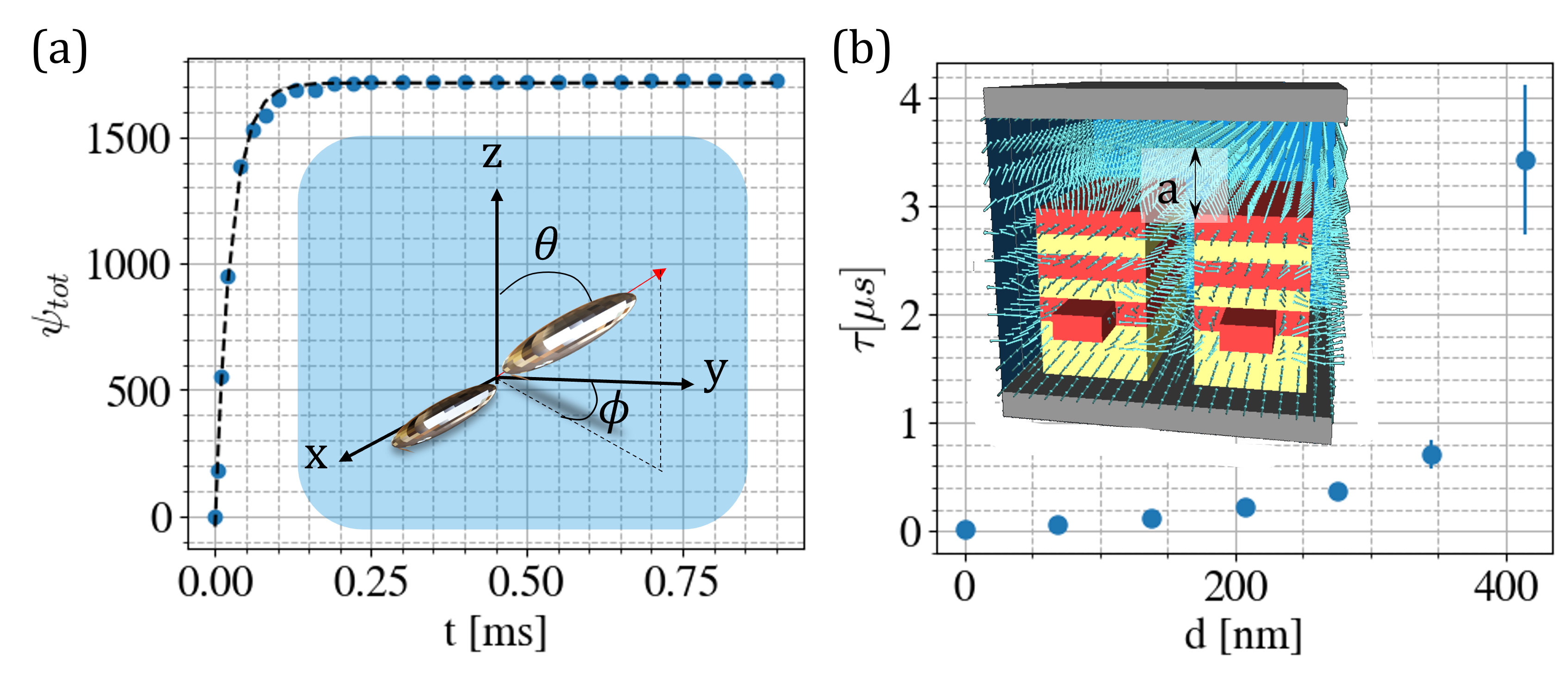}
  \caption{Shintech LC simulations. (a) Integrated angular deviation over time $\psi(t)_{tot}$ between the electrodes with an applied 5 V at $t=0$ ms and exponential fit (black, dashed). Inset shows the azimuthal and polar angles of the LC, $\theta(x,y,z,t)$ and $\phi(x,y,z,t)$. (b) Fitted rise time as a function of the distance from the electrodes $a$. Inset is an image from the Shintech simulation showing the nano-rod structures and LC orientations when applying 4.5 V across the electrodes. $a$ is the distance from the top of the electrodes.}\label{fig-sims}
\end{figure}

To evaluate the rotational dynamics of LC molecules, we conducted simulations using Shintech LCDMaster3D. Fig. \ref{fig-sims}(a) shows the simulated integrated angular deviation $\psi(t)_{tot}$ over the simulation time, and the exponential fit used to extract the rise time. The angular deviation was integrated over the area between the rods, where the electric field attains a high magnitude, as obtained from the COMSOL simulations, and can be seen in Fig. \ref{fig2}(c). \\
Additionally, we explored how this rise time varies with respect to the vertical distance from the nanorods, denoted as $a$ and defined in Fig. \ref{fig-sims}(a)(inset). Our findings indicate an exponential increase in the LC rotation time with increasing distance from the nano-rods. Fig. \ref{fig-sims}(b) shows the rise time as a function of $a$, where this exponential relation is evident. The rise time at $a$ = 414 nm above the nanorods is 3.4 $\pm $0.6 ms, surpassing the value at $a$ = 0 nm by more than 200-fold. 

\end{appendices}


\bibliography{sn-bibliography}

\end{document}